\documentclass[aps,prd,twocolumn]{revtex4}

\usepackage{subfigure}
\usepackage{graphicx}
\usepackage{amssymb}
\usepackage{color}
\usepackage{times}
\usepackage{xspace}

\newcommand{\et}{{E}_{\scriptscriptstyle\rm T}}
\newcommand{\met}{\mbox{$\protect \raisebox{.3ex}{$\not$}\et$}}

\newcommand{\wenu}{W \rightarrow e \nu}
\newcommand{\zee}{Z \rightarrow e^{+}e^{-}}

\newcommand{\ppbar}{p\bar{p}}
\newcommand{\yW}{y_{W}}

\newcommand{\ifb}{\ensuremath{\rm fb^{-1}}\xspace}
\newcommand{\MeV}{\ensuremath{\mathrm{Me\kern-0.1em V}}\xspace}
\newcommand{\GeV}{\ensuremath{\mathrm{Ge\kern-0.1em V}}\xspace}
\newcommand{\GeVc}{\ensuremath{\mathrm{Ge\kern-0.1em V}}\xspace}
\newcommand{\GeVcc}{\ensuremath{\mathrm{Ge\kern-0.1em V}}\xspace}
\newcommand{\TeV}{\ensuremath{\mathrm{Te\kern-0.1em V}}\xspace}

\def\Z0{{\em $Z^{0}$\/}}

\begin{document}




\title{A new analysis technique to measure the $W$ Production Charge Asymmetry at the Tevatron}
\author {Arie Bodek, Yeonsei Chung, Bo-Young Han, and Kevin McFarland }
\affiliation{Department of Physics and Astronomy, University of Rochester, Rochester, New York 14627}
\author {Eva Halkiadakis}
\affiliation{Department of Physics and Astronomy, Rutgers University, Piscataway, New Jersey 08855}
\date{\today}


\begin{abstract}
  
The charge asymmetry of $W$ bosons produced in $p\bar{p}$ collisions
at $\sqrt{s} = 1.96$ \TeV is sensitive to the ratio of $d$ and $u$
quark distributions in the range of $ x > 0.002$ at $Q^2 \approx
M_{W}^{2}$.  We propose an analysis technique to directly measure $W$
production charge asymmetry from $\wenu$ events at the Tevatron and
show the feasibility for this method using Monte Carlo simulations.

\end{abstract}

\maketitle

\section{Introduction}
\begin{figure}
  \begin{center}
  \includegraphics[width=7.5cm]{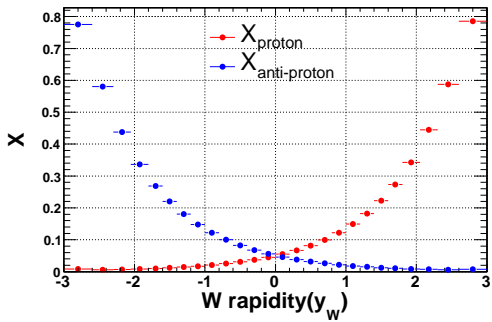}
    \caption{The x values of the quark for $W$ production at the Tevatron.     
   \label{fig:xVSwrap} }
  \end{center}
\end{figure}

The differential cross section for $W$ boson production in $\ppbar$
as a function of $W$ rapidity is
\begin{eqnarray}
\frac{d\sigma^{\pm}}{dy_{W}} & = & \frac{2\pi}{3} \frac{G_{F}}{\sqrt{2}} 
\sum_{q\bar{q}} \left| V_{q\bar{q}} \right|^2 \left[ q(x_{p})\bar{q}(x_{\bar{p}}) + \bar{q}(x_{p}){q}(x_{\bar{p}}) \right], \nonumber \\
\end{eqnarray}
where $x_{p}$ ($x_{\bar{p}}$) is the fraction of the proton
(anti-proton) momentum carried by the struck quark, $q$ and $\bar{q}$
are the quark and anti-quark parton distribution functions, and $\yW$ is the
rapidity of the $W$ boson.  The x values of the quark in the proton and antiquark 
in the antiproton are related to the rapidity, $y$, of the W  boson via the equation 
$x_{p,\bar{p}} = M_W/\sqrt{s}e^{\pm y_W}$ as shown in Fig~\ref{fig:xVSwrap}. 
Here $\sqrt{s}$ is the center of mass energy and $M_W$ is the mass of the W boson.
 
\begin{figure}
  \begin{center}
    \subfigure[]{\label{fig:wrap}\includegraphics[width=4.2cm,]{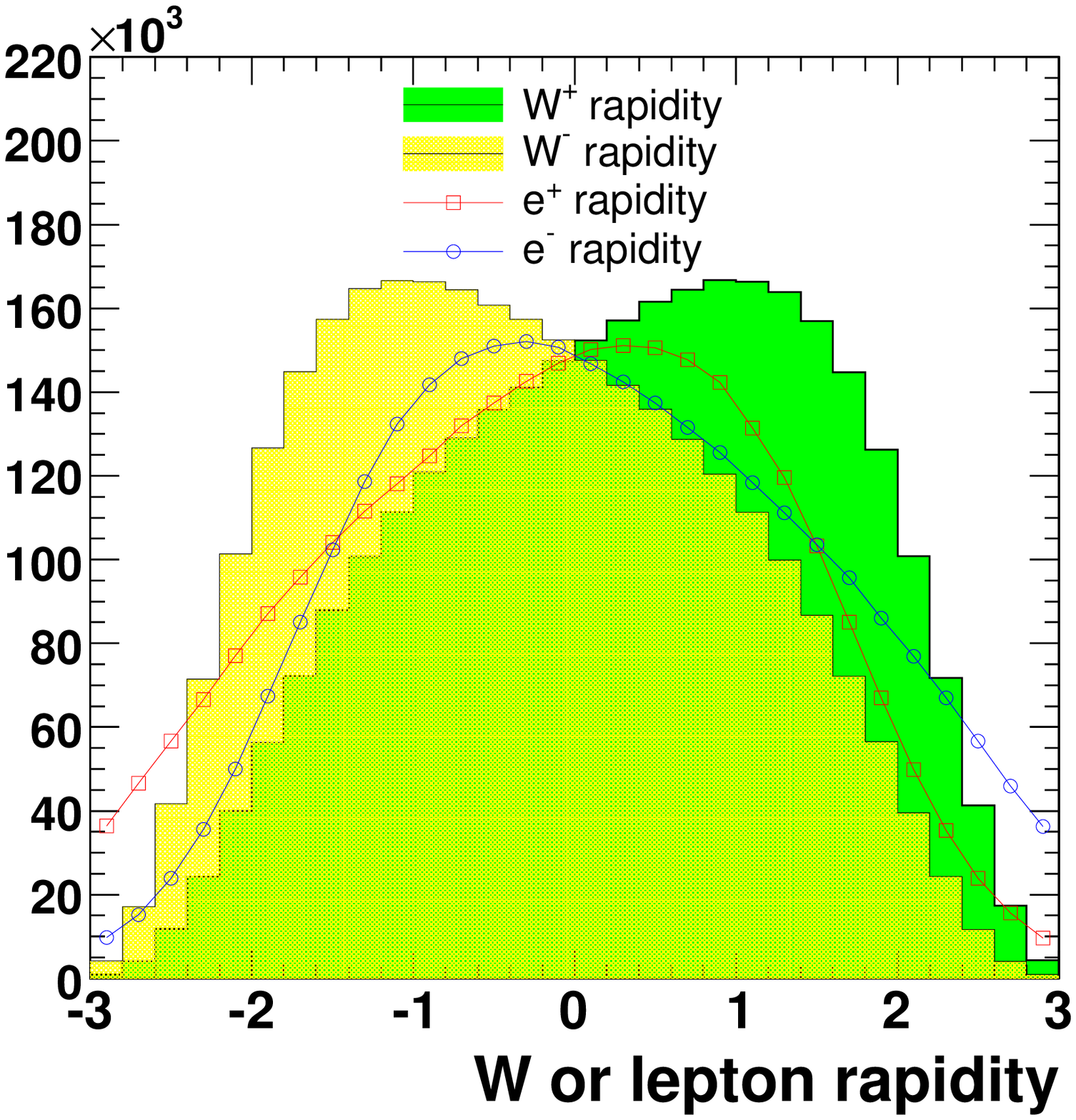}}
    \subfigure[]{\label{fig:wasym}\includegraphics[width=4.2cm,]{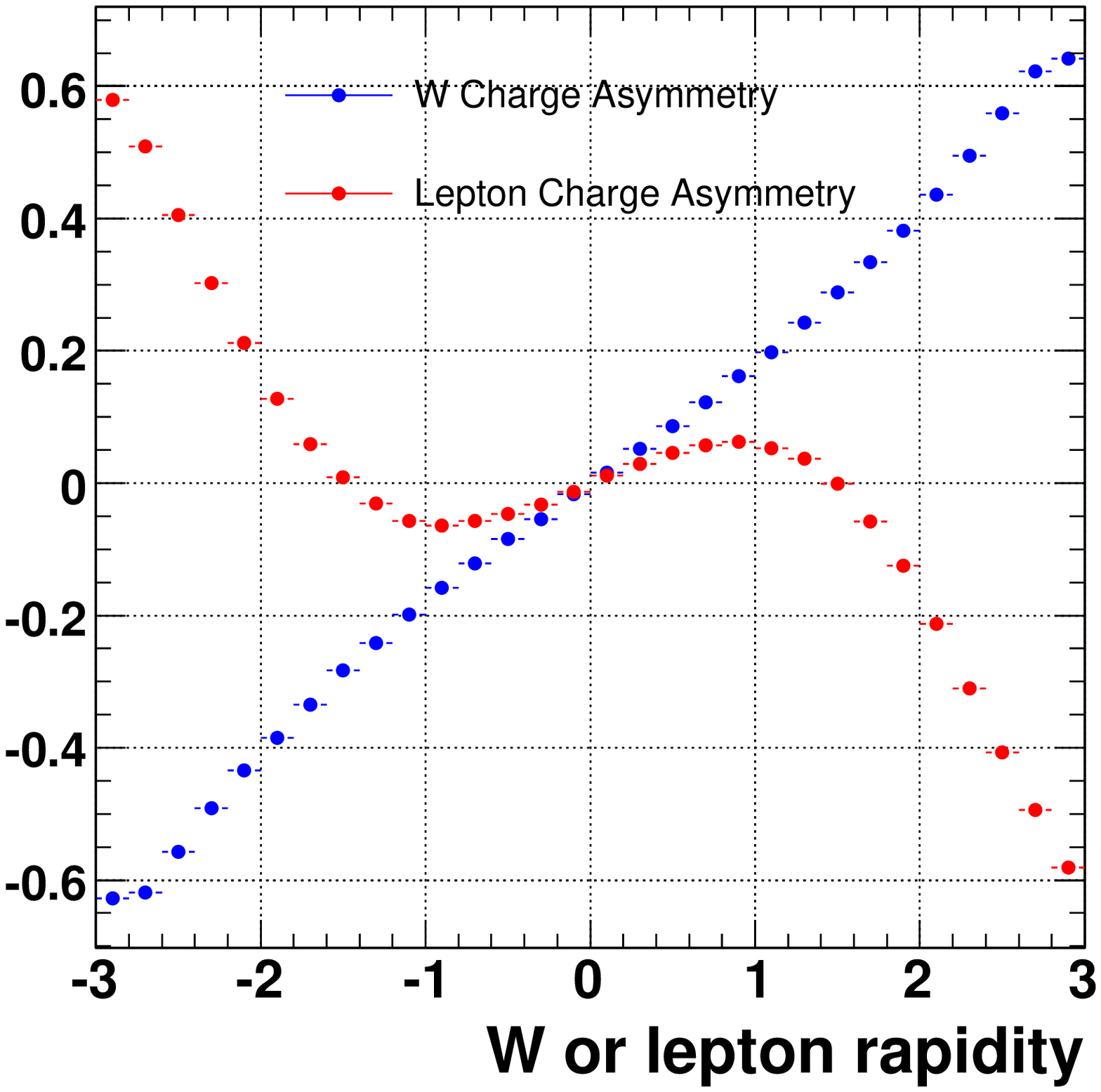}}
    \caption{~(a)~The $W$ boson and lepton rapidity distributions in $\ppbar$ collisions.
             ~(b)~The charge asymmetry for $W$ production and the decay lepton.}  \end{center}
\end{figure}

Since $W^+ (W^-)$ bosons are produced in $\ppbar$ 
collisions primarily by the annihilation of $u (d)$ quarks in the proton and $\bar{d} 
(\bar{u})$ quarks in the anti-proton, and since $u(x_{p})=\bar{u}(x_{\bar{p}})$ and
$d(x_{p})=\bar{d}(x_{\bar{p}})$ by CPT symmetry, the differential cross
sections for $W^{\pm}$ are approximately
\begin{eqnarray}
\frac{d\sigma^{+}}{dy_{W}} & \approx & \frac{2\pi}{3} \frac{G_{F}}{\sqrt{2}}
\left[ u(x_{p})\bar{d}(x_{\bar{p}}) \right], \label{eq:dsdyp}\\
\frac{d\sigma^{-}}{dy_{W}} & \approx & \frac{2\pi}{3} \frac{G_{F}}{\sqrt{2}}
\left[ d(x_{p})\bar{u}(x_{\bar{p}}) \right]. \label{eq:dsdym}
\end{eqnarray}
Since the $u$ quark tends to carry a larger fraction of the proton's momentum
than the $d$ quark on average, the $W^+ (W^-)$ is boosted in the proton
(anti-proton) direction as shown in Fig.~\ref{fig:wrap}.  
The $W$ production charge asymmetry, $A(y_W)$, in the leading-order
parton model is therefore
\begin{eqnarray}
A(y_W) & = & \frac{d\sigma^{+}/dy_{W} - d\sigma^{-}/dy_{W}}{d\sigma^{+}/dy_{W} + d\sigma^{-}/dy_{W}} \nonumber \\
      & \approx & \frac{u(x_p)\bar{d}(x_{\bar{p}}) - d(x_p)\bar{u}(x_{\bar{p}})}{u(x_p)\bar{d}(x_{\bar{p}}) + d(x_p)\bar{u}(x_{\bar{p}})} \nonumber \\
      & = & \frac{R_{du}(x_{\bar{p}}) - R_{du}(x_{p})}{R_{du}(x_{\bar{p}}) + R_{du}(x_{p})},
\label{eq:asym}
\end{eqnarray}
where we use Eq.~\ref{eq:dsdyp} and Eq.~\ref{eq:dsdym} and introduce the ratio 
$R_{du} = \frac{d(x)}{u(x)}$. 
As we see in Eq.~\ref{eq:asym}, there is a direct correlation between
the $W$ production charge asymmetry and the $d/u$ ratio.  A precise
measurement of the $W$ production charge asymmetry serves as a
valuable constraint on the $u$ and $d$ quark momentum
distributions~\cite{text1}.  

$W$ production at hadron colliders is identified through the process 
$p + \bar{p} \rightarrow W^{\pm}; 
W^{\pm} \rightarrow \ell^{\pm} + \nu $.
Since the $W$ decay involves a neutrino
whose longitudinal momentum is experimentally undetermined, the charge
asymmetry previously has been constrained by the measured charge
asymmetry of the decay leptons and as a function of the lepton
pseudo-rapidity~\cite{run1,run2,d0prelim}.  However, as shown in
Fig.~\ref{fig:wasym}, there is a ``turn-over" in the
lepton charge asymmetry due to a convolution of the $W$ production
charge asymmetry and the $W$ $V-A$ decay. 
This convolution means leptons from a single pseudorapidity come from
a range of $W$ rapdities and thus a range of parton $x$ values.  Thus,
the measured lepton asymmetry is more complicated to interpret in
terms of quark distributions, and we expect the direct measurement of
the asymmetry of the $W^{\pm}$ rapidity distribution to be a more
sensitive probe of the differences between $u$ and $d$ quarks.

In this paper, we propose a new analysis technique which resolves the
kinematic ambiguity of the longitudinal momentum of the neutrino to
directly reconstruct the $W^{\pm}$ rapidity. We describe the details
of our new analysis technique and outline the sources of systematic
uncertainty of this measurement.  Our studies are performed in the
$W^{\pm} \rightarrow e^{\pm}\nu$ channel produced in $\ppbar$
collisions at the Tevatron.  We use a realistic Monte Carlo simulation (MC@NLO)
and include the effects of higher-order QCD corrections~\cite{mcatnlo}.

\section{Analysis Technique}
The $W$ decay to leptons, in our case $W^{\pm} \rightarrow
e^{\pm}\nu$, involves a neutrino whose longitudinal momentum cannot
be experimentally determined.  However, we can determine the longitudinal
momentum by constraining the $W$ mass in Eq.~\ref{eq:Mw}, which results 
in a two-fold ambiguity. This ambiguity can be partly resolved on a 
statistical basis from the known $V-A$ (vector-axial vector) decay 
distribution using the center-of-mass decay angle between the electron 
and the proton, $\theta^*$, and from the $W^+$ and $W^-$ production 
cross-sections as a function of $W$ rapidity, $d\sigma^{\pm}/dy_{W}$.
The $W$ mass constraint is
\begin{equation}
  M^2_W = (E_l + E_\nu)^2 - (\vec{P_l} + \vec{P_\nu})^2,  \label{eq:Mw}
\end{equation}
where the $W$ mass, $M_{W}$, 
is contrained to its experimentally measured
value~\cite{wmass-cdfrun2,pdg}.
Events which cannot satisfy the $W$ mass constraint (and which get
imaginary values of the neutrino $z$-momentum) are due to a
mis-reconstruction of the neutrino (missing) transverse energy, $\met$~\cite{coord}. 
Therefore, in such cases, we re-scale the $\met$ to the value which makes 
the imaginary part to be zero. This new $\met$ is then used to correct 
the $y_W$ for the event. 

\begin{figure}
  \begin{center}
    \subfigure[]{\label{fig:Wpl_pos}\includegraphics[width=4.2cm]{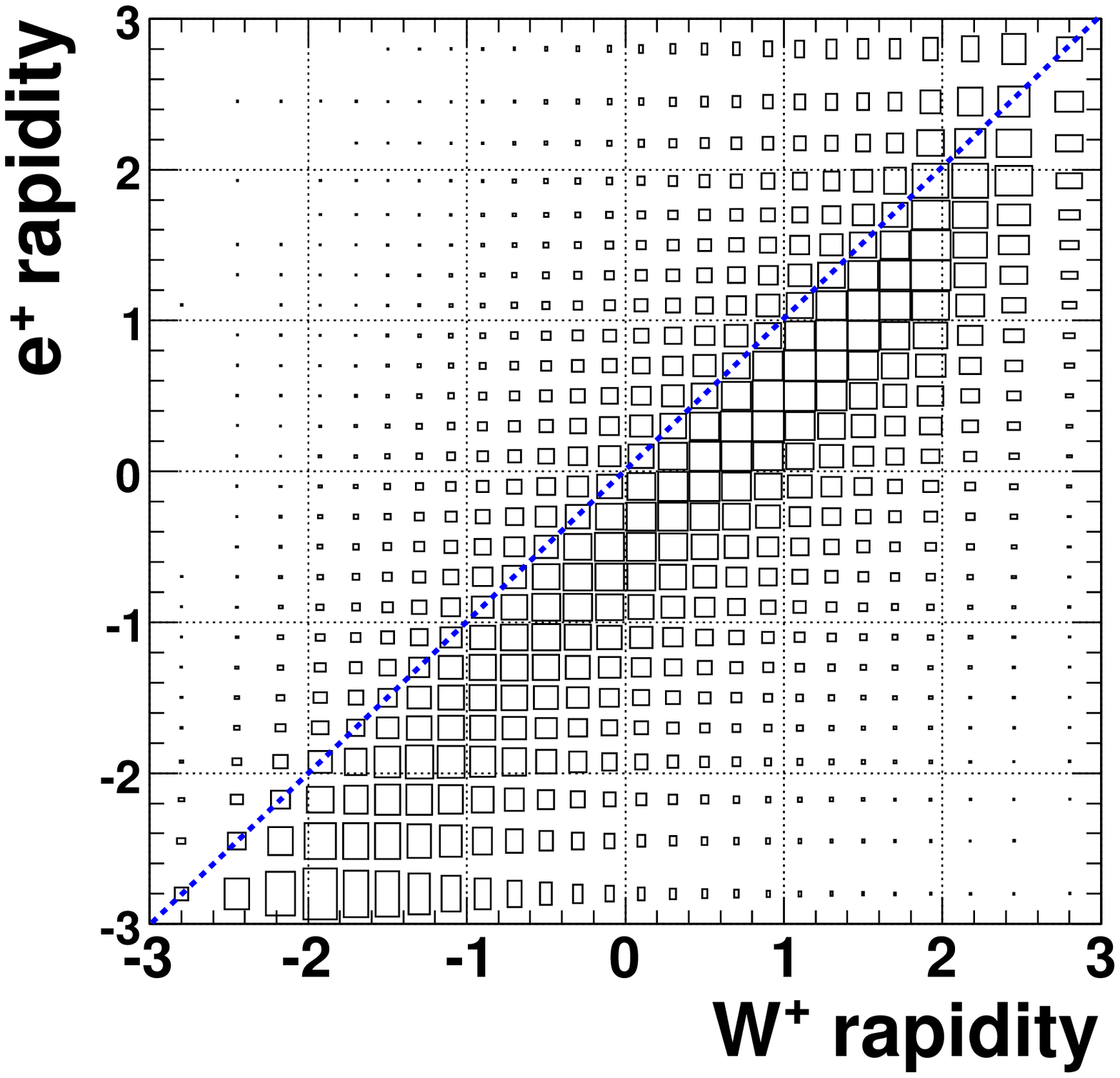}}
    \subfigure[]{\label{fig:Wmi_ele}\includegraphics[width=4.2cm]{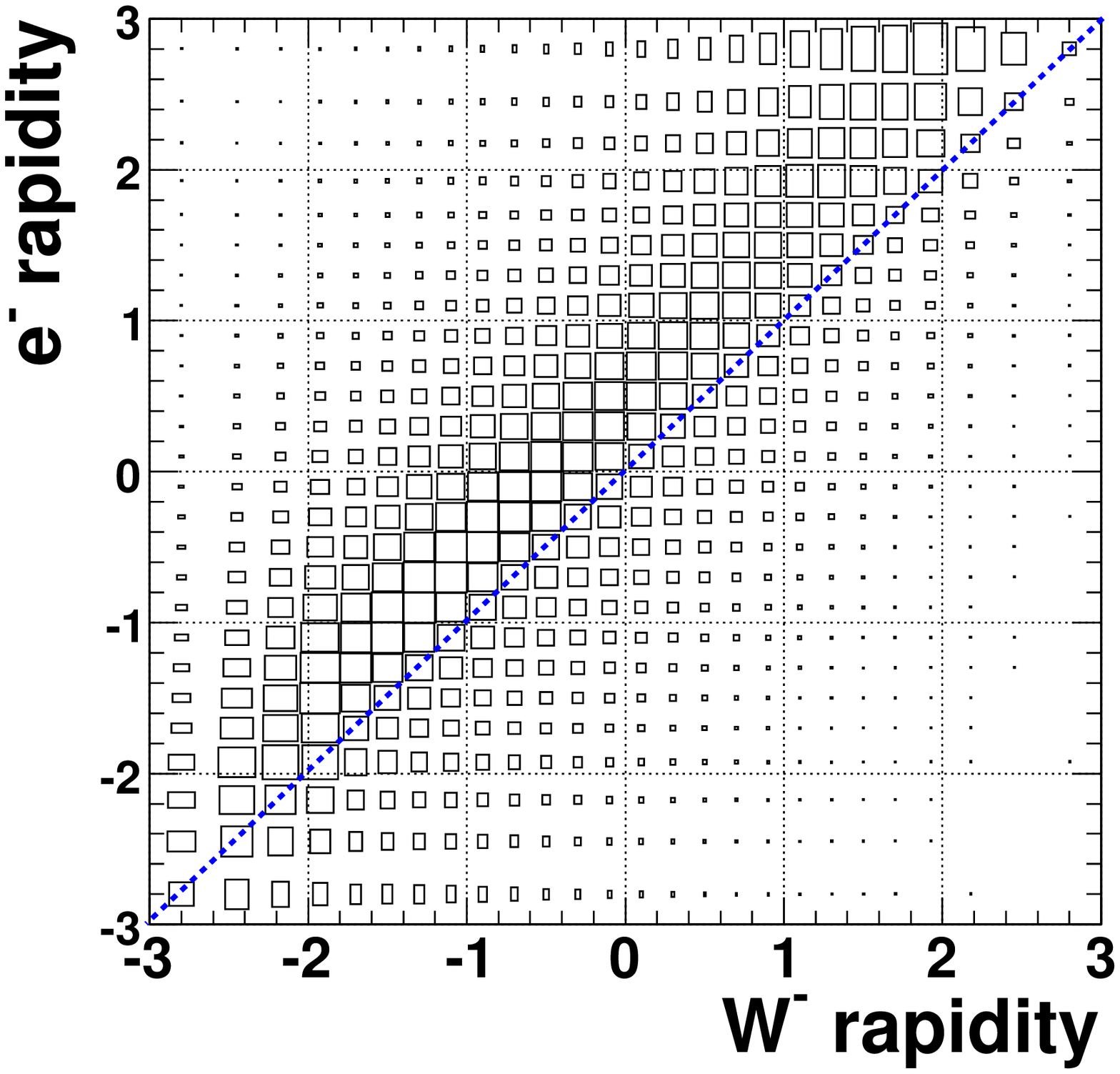}}
    \caption{~(a)~The positively charged $W$ boson and lepton rapidity distribution.
             ~(b)~The negatively charged $W$ boson and lepton rapidity distribution.}
  \end{center}
\end{figure}

The leading order $W$ boson production mechanism in $p\bar{p}$
collisions results in the $W$ boson being polarized in the $\bar{p}$
direction by means of the $V-A$ structure of the weak interaction. The
$V-A$ structure means that the weak current couples only to left-handed
$u$ and $d$ quarks (or to right-handed $\bar{u}$ and $\bar{d}$
quarks). For ultra-relativistic quarks, where helicity and chirality
are approximately equivalent, this results in full polarization of the
produced $W$ bosons in the direction of the beam. The $W$ leptonic
decay process also couples only to left-handed $e^{-}$ and
right-handed $\bar{\nu}$ (or right-handed $e^{+}$ and left-handed
$\nu$). The conservation of angular momentum favors a decay with the
final state lepton (neutrino or electron) at a small angle with
respect to the initial state quark direction (and a similar small
angle between the initial state anti-quark and final anti-lepton). The
systematic shift in lepton pseudo-rapidity with respect to $y_W$
depending on the charge of the final state lepton is illustrated in
Fig.~\ref{fig:Wpl_pos} and ~\ref{fig:Wmi_ele}, which shows the lepton 
pseudo-rapidity vs. $W$ rapidity for the different charges. This effect 
also explains the discrepancy at high rapidity between the lepton 
charge asymmetry and the $W$ charge asymmetry as illustrated in
Fig.~\ref{fig:wasym}.  The $V-A$ bias in the $W$ decay
angle causes leptons at high rapidity to originate primarily from $W$
bosons produced in the opposite hemisphere.

$W^\pm$ bosons at the Tevatron are primarily produced from the valence
quarks in the proton and the anti-quarks in the anti-proton and rarely
from sea quarks simply because $W$ production requires at least one
moderately high $x$ parton to be involved in the
collision.  At very large forward or backward rapidities
where one very high $x$ parton must participate in the production, the
production probability from the sea quarks nearly
vanishes. Understanding of the sea quark contribution is important to
exactly know the decay angle distributions from the $V-A$ structure
because $W$ production by sea anti-quarks will result in the opposite
$W$ polarization from valence quark production.

\begin{figure}
  \begin{center}
    \subfigure[]{\label{fig:costheta}\includegraphics[width=4.2cm]{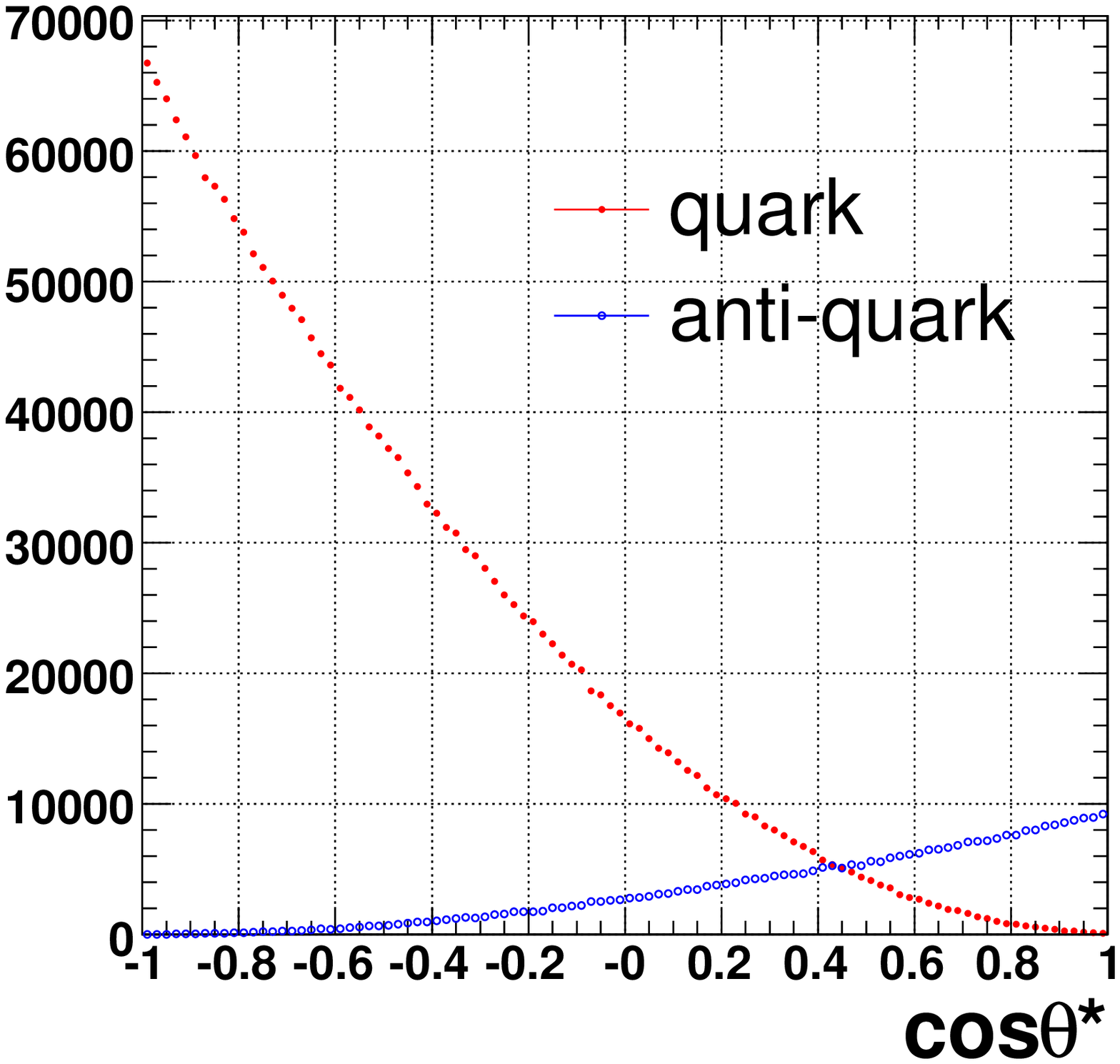}}
    \subfigure[]{\label{fig:qoverqbar}\includegraphics[width=4.2cm]{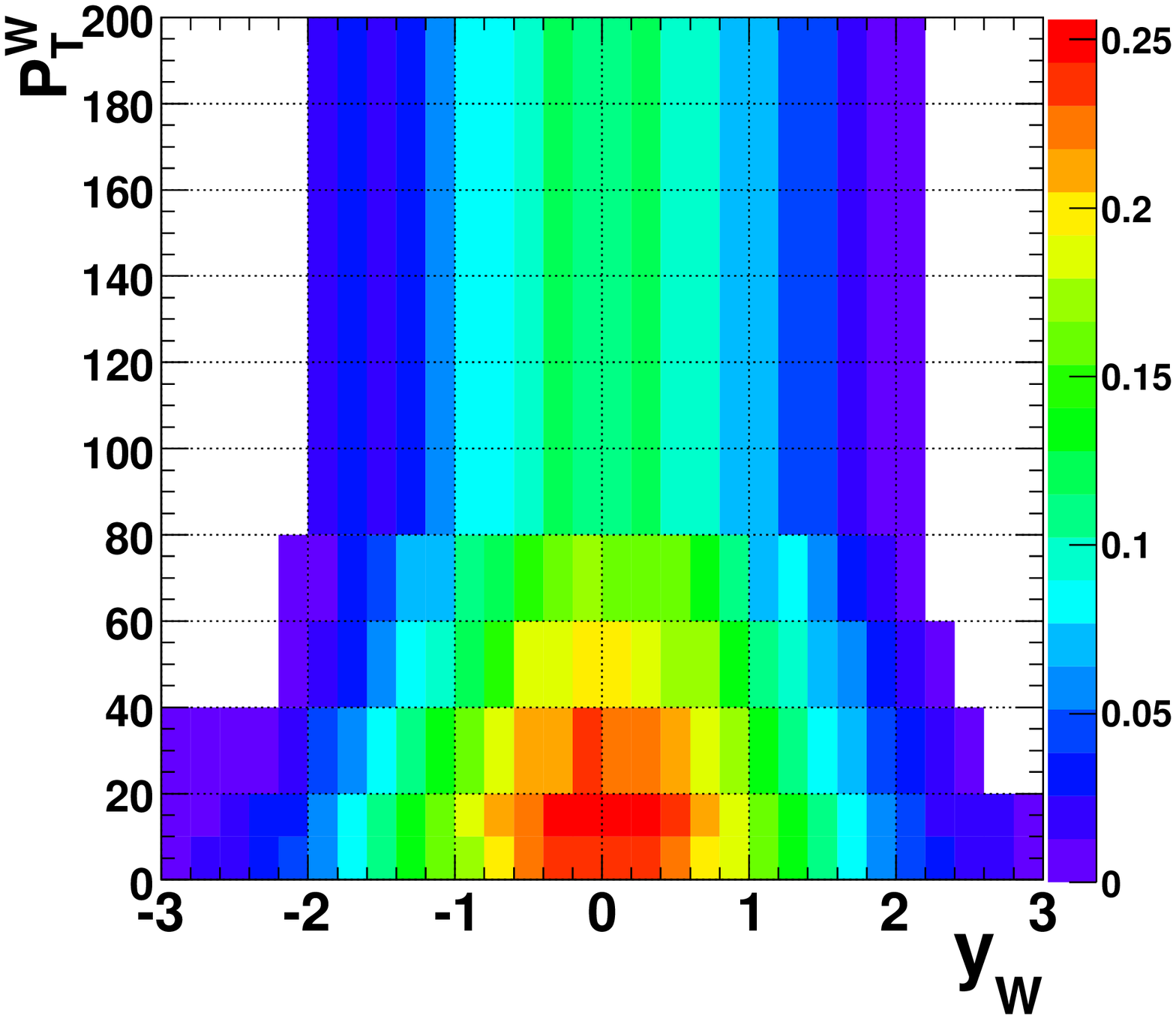}}    \caption{~(a)~The cos$\theta^*$ distributions of $e^+$ in the $W^+$
        rest frame, averaged over all produced $W^+$.
        The curve labeled ``quark'' shows the case when a quark from
        the proton and anti-quark from the anti-proton form the $W^+$.
        The curve labeled ``anti-quark'' shows the opposite case, when
        an anti-quark from the proton and a quark from the anti-proton
        form the $W^+$.
             ~(b)~The dependence of the ratio of ``anti-quark''
      ($\bar{q}$) and ``quark'' ($q$) contributions to the overall
      $W$ decay angle distribution, $Q(y_W,p_T^W)$, as a function of $W$
        rapidity and $p_T$ of the $W$.}
  \end{center}
\end{figure}

We use a Monte Carlo simulation with NLO QCD
corrections~\cite{mcatnlo} to determine the production probability
with sea quarks by identifying initiating quarks as a function of
$y_W$.  We verify the expected angular distribution of $(1 \pm
cos\theta^*)^2$ from production of $W^\pm$ with quarks in the proton
and the opposite distribution with anti-quarks in the proton.  For
example, in Fig.~\ref{fig:costheta}, we show the cos$\theta^*$
distributions of $e^+$ in the $W^+$ rest frame for the case when a
quark from the proton and an anti-quark from the anti-proton form the
$W^+$ (labeled ``quark'') and the case when an anti-quark from the
proton and a quark from the anti-proton form the $W^+$ (labeled
``anti-quark''). The ratio of quark (proton) and anti-quark (proton)
induced $W$ production, therefore, determines the angular decay
distribution.  In the simulation, we measure the fraction of quark and
anti-quark contributions, and parameterize the angular distributions
for $y_{W}$ and the $W$ transverse momentum, $p_{T}^{W}$.  We find an
empirical
functional form that fits the data,
\begin{equation}
P_{\pm}(cos\theta^*,y_W,p_{T}^{W}) = (1 \mp cos\theta^*)^2 + Q(y_W,p_{T}^{W})(1 \pm cos\theta^*)^2, \label{eq:prob}
\end{equation}
\begin{equation}
Q(y_W,p_T^W) = f(p_T^W)e^{-[g(p_T^W)*{y_W}^2 + 0.05*|{y_W}^3|]}. \label{eq:qfun}
\end{equation}
The parameters $f(P_T^W)$ and $g(P_T^W)$ are
\begin{eqnarray}
f(P_T^W) & = &  0.2811{\cal L}(P_T^W,\mu = 21.7 \GeV , \sigma = 9.458 \GeV ) \nonumber \\
         & & + 0.2185e^{(-0.04433 \GeV ^{-1}P_T^W)}, \nonumber \\
g(P_T^W) & = &0.2085 + 0.0074\GeV^{-1}P_T^W \nonumber \\
& & - 5.051\times10^{-5}\GeV^{-2}{P_T^W}^2 \nonumber \\
& & + 1.180\times10^{-7}\GeV^{-3}{P_T^W}^3,
\end{eqnarray} 
\noindent
where ${\cal L}(x,\mu,\sigma)$ is the Landau distribution with most
probable value $\mu$ and the RMS $\sigma$.  The first term of
Eq.~\ref{eq:prob} corresponds to the contribution from quarks in the
proton and the second term from anti-quarks in the proton.  The
parameterization, $Q(y_W,p_T^W)$, the ratio of the two angular
distributions as a function of the $W$ rapidity and $p_T^W$, is
obtained from the fit to the distribution in Fig.~\ref{fig:qoverqbar}.

A second relevant factor in the selection among the two $W$ rapidity
solutions is the $W$ differential cross-section as a function of
$y_W$, $d\sigma^{\pm}/dy_{W}$. The $W$ boson production decreases
sharply beyond $|y_W|>2$ because of the scarcity of high $x$
quarks. For instance, if one of the two possible solutions falls in
the central region of rapidity and the other has $|y_W|>2$, the former
should receive more weight as the latter is very unlikely to be
produced.

The information used to select among the two solutions can be
represented by a weighting factor for each rapidity solution and charge,
$w^{\pm}_{1,2}$, can be represented as

\begin{eqnarray}
w^{\pm}_{1,2} & = \frac{P_{\pm}(cos\theta^*_{1,2},y_{1,2},p_T^W) \sigma^{\pm}(y_{1,2})}{P_{\pm}(cos\theta^*_{1},y_{1},p_T^W) \sigma^{\pm}(y_{1}) + P_{\pm}(cos\theta^*_{2},y_{2},p_T^W) \sigma^{\pm}(y_{2})}, 
\label{eq:wt}
\end{eqnarray}
where the $\pm$ signs indicate the $W$ boson charge and indices of 1,
2 are for the two $W$ rapidity solutions.  

\begin{figure}
\centering
  \begin{center}
    \subfigure[]{\label{fig:smearing}\includegraphics[width=8.2cm,]{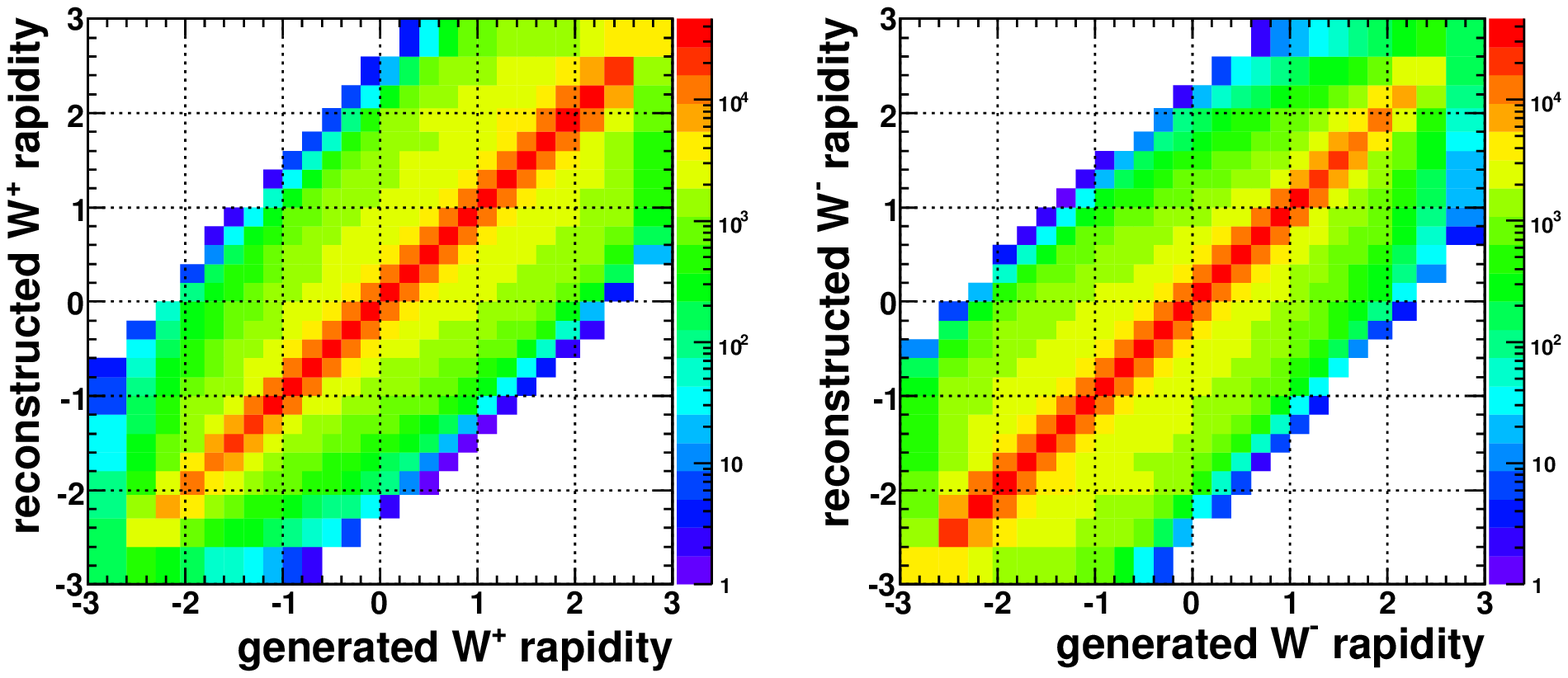}}

    \subfigure[]{\label{fig:accept}\includegraphics[width=7.0cm,height=4.5cm]{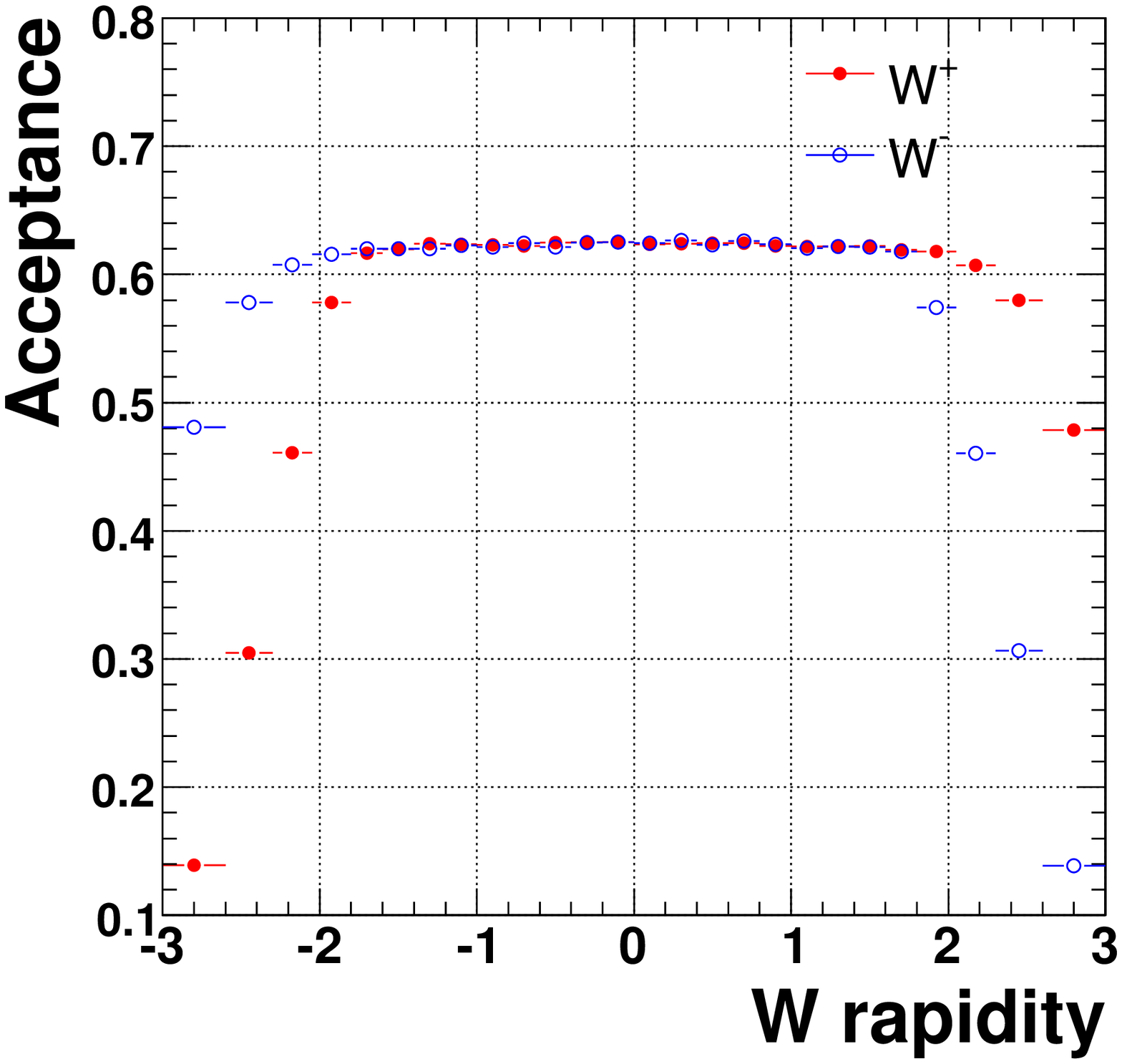}}
    \caption{~(a)~ Comparison of the simulated $W$ boson rapidity with the
        generated $W$ rapidity. The simulated rapidity is reconstructed by using
        the weighting factor and the smearing effect is considered.
             ~(b)~ Acceptance distribution for $\wenu$ events as a function of $\yW$.
        The solid (open) circles represent $W^{+} (W^{-})$ events.}
  \end{center}
\end{figure}

In our analysis, we include kinematic cuts for detecting charged
leptons.  For $\wenu$ event selection, we apply $|\eta_{e}^{lab}| <
2.8$, $\et^{e} > 25 \GeV$, and $\met > 25 \GeV$.  We also consider a
multiplicative correction factor for the detector acceptance and event
migration from smearing effects as shown in Figs.~\ref{fig:smearing}
and ~\ref{fig:accept}.  In order to study smearing effects, we use the
fact that the energy resolutions in the electromagnetic calorimeter of
the Collider Dectector at Fermilab (CDF) are $14\%/\sqrt{\et}$
(central calorimeter) and $16\%/\sqrt{E} \oplus 1\%$ (the end plug
calorimeter) and in the hadronic calorimeter are $75\%/\sqrt{E}$
(central) and $80\%/\sqrt{E} \oplus 5\%$ (the end plug)~\cite{TDR}.
We randomly smear the electron and recoil hadronic energies in simulated events
with a Gaussian distribution modeling their uncertainties prior to making 
the selection above.  
The correction factors
are determined using a Monte Carlo program which includes both a model
of the process under study as well as a simulation of the measuring
apparatus.  
In Eq.~\ref{eq:wt}, the weighting factor depends primarily
on the $W^+$ and $W^-$ cross-sections, but does have some weak
dependence on the assumed $W$ charge asymmetry, and thus the
correction factors can be biased by computing the factors with
different Monte Carlo models.  Therefore, this method requires us to
iterate the procedure to eliminate our measurement's dependence on the
input asymmetry.  In order to confirm our analysis technique and take
into account the bias from physics input variables (such as the charge
asymmetry itself, the total differential cross-section and the angular
distribution) we have studied the $W$ charge asymmetry measurement
with different Monte Carlo models and evaluated systematic
uncertainties, which are described in the next section.

\section{Systematic Uncertainties}
We consider potentially significant sources of systematic uncertainty on
the $W$ charge asymmetry measurement from the assumed parton
distributions, the detector resolutions and misidentifications and
backgrounds.  Input PDFs are used to determine the parameters of the
weighting factor, and may affect the final result.  The detector
resolutions affect the $W$ rapidity reconstruction due to
uncertainties in the calorimeter energy scale and its energy
resolution, and the missing transverse energy scale also has a
significant uncertainty from the $W$ boson recoil energy scale.
Finally, the detector may misidentify the charge, especially from
leptons at high $\left| y_W\right| $, and there are backgrounds to
$W\to e\nu$ at the Tevatron.

The uncertainties on the weighting factor (Eq.~\ref{eq:wt}) arise
from uncertainties on the momentum distribution of quarks and gluons
in the proton modeled with the PDF sets used. The choice of PDF set
has an effect on the shape of the $d\sigma^{\pm}/dy_{W}$ distribution
as well as on the ratio of quark and anti-quark in the angular decay
distribution. We use the CTEQ6 error PDF sets~\cite{cteq6m} and
re-determine the $d\sigma^{\pm}/dy_{W}$ production cross section and
the angular distribution of $(1 \pm cos\theta^{*})^{2}$ for each error
PDF set. We evaluate the uncertainty on the $W$ charge asymmetry by
checking the deviation of the asymmetry values based on each
calculation from the central value obtained using the best-fitted PDF set.

We also consider several experimental sources of systematic
uncertainty. The scale and resolution of the electromagnetic
calorimeter energy and the missing transverse energy ($\met$)
can change the $W$ rapidity and thus the asymmetry measurement. 
We use the energy uncertainties measured in ~\cite{wzxsec}, where the 
electromagnetic (EM) calorimeter energy scale and resolution was tuned in 
the simulation so as to fit to data. The uncertainty on the energy scale 
and resolution was measured to be $0.3\%$ (scale), $1.5\%$ (resolution)
for central electron and $0.6\%$ (scale), $1.1\%$ (resolution) for plug
electron. These values correspond to a 3 $\sigma$ variation.
The asymmetry uncertainties are estimated as the changes in the measured
asymmetry when the energy scale and resolution are changed between its 
default and the $\pm 3 \sigma$ value.
The missing transverse energy ($\met$) in our $\wenu$ sample is 
determined by the assumption that the vector sum of all transverse energy 
should be zero. Since hadronic transverse energy is due to the $W$ boson 
recoil energy, we consider the transverse recoil energy, which is affected 
by multiple interactions in the event. The uncertainty on the transverse 
recoil energy scale is $2\%$ (3$\sigma$)~\cite{wzxsec}.
The charge misidentification rate and background estimates are crucial
for the charge asymmetry measurement since both can directly change
the measurement.  We estimate these uncertainties using the charge
fake rates (CFR) and background fractions (BKG) from the previous $W$
lepton charge asymmetry result from CDF~\cite{run2}. 
The charge fake rate is about 0.01 for $|\eta| < 1.5$ and
0.04 for $|\eta| > 2.0$. The upper bound on the background fraction is 
$2\%$ for $|\eta| < 1.0$ and increases to about $15\%$ for $|\eta| > 2.0$.
We also investigate sources of any charge bias and $\eta$ dependence
in the kinematic and geometrical acceptance of the event.
An uncorrected acceptance shift of 3$\%$ central and forward electrons and 
5$\%$ far forward electrons ($|\eta| > 2.4$) based on measurements of $\zee$ 
data~\cite{dzdy} are taken to address the effects of systematic on $W$ charge 
asymmetry measurement.

Table~\ref{tab:syst_err} summarizes the systematic uncertainties on
the $W$ boson production charge asymmetry for rapidities $|y_W| < 3.0$.  
We compare the expected statistical uncertainty obtained
by assuming an analysis using an integrated luminosity of 1 $\rm{fb^{-1}}$,
where we also extrapolate the expected statistical uncertainty from the number of events
from the previous $W$ lepton charge asymmetry result of CDF with 0.2$\ifb$~\cite{run2}. 

\begin{table}[tbp]
\caption{Systematic and statistical uncertainties assuming an integrated luminosity of $1fb^{-1}$ for the $W$ production charge asymmetry in $W$ rapidity bins.\label{tab:syst_err}}
\begin{ruledtabular}
\begin{tabular}{c c c c c c c c c}
                  & \multicolumn{8}{c}{ $\Delta A$ ($\times 10^{-3}$) } \\
 $W$ rapidity     & PDF & EM & Recoil & CFR & BKG & Accep & Syst. & Stat. \\ \hline
0.0$< |y_{W}| <$0.2  & 0.1 & 0.1 & 0.1 & 0.1 & 0.2 & 0.0 & 0.3 & 6.3  \\
0.2$< |y_{W}| <$0.4  & 0.4 & 0.1 & 0.1 & 0.1 & 0.3 & 0.3 & 0.6 & 6.4  \\
0.4$< |y_{W}| <$0.6  & 0.5 & 0.2 & 0.2 & 0.2 & 0.5 & 0.9 & 1.3 & 6.4  \\
0.6$< |y_{W}| <$0.8  & 0.6 & 0.3 & 0.3 & 0.2 & 0.5 & 1.6 & 1.9 & 6.5  \\
0.8$< |y_{W}| <$1.0  & 0.6 & 0.2 & 0.3 & 0.2 & 0.6 & 2.2 & 2.4 & 6.6  \\
1.0$< |y_{W}| <$1.2  & 0.6 & 0.2 & 0.2 & 0.3 & 0.6 & 2.8 & 2.9 & 6.7  \\
1.2$< |y_{W}| <$1.4  & 0.7 & 0.4 & 0.1 & 0.7 & 0.5 & 3.0 & 3.2 & 7.0  \\
1.4$< |y_{W}| <$1.6  & 0.7 & 0.3 & 0.2 & 1.4 & 0.2 & 2.9 & 3.3 & 7.5  \\
1.6$< |y_{W}| <$1.8  & 0.7 & 0.5 & 0.5 & 2.9 & 0.2 & 2.8 & 4.1 & 8.1  \\
1.8$< |y_{W}| <$2.05 & 0.6 & 0.5 & 0.5 & 4.4 & 0.8 & 2.9 & 5.4 & 8.0  \\
2.05$< |y_{W}| <$2.3 & 0.7 & 0.4 & 0.4 & 6.3 & 1.5 & 3.4 & 7.4 & 9.6  \\
2.3$< |y_{W}| <$2.6  & 0.7 & 0.7 & 0.3 & 6.1 & 1.3 & 4.9 & 8.0 & 11.7  \\
2.6$< |y_{W}| <$3.0  & 2.9 & 1.0 & 0.9 & 1.1 & 0.3 & 5.0 & 6.0 & 17.0  \\
\end{tabular}
\end{ruledtabular}
\end{table}

\section{Results}

We compare the expected statistical uncertainties in 1.0 fb$^{-1}$ of
data at the Tevatron with the uncertainties coming from parton
distribution functions (PDFs) using CTEQ6M in
Fig.~\ref{fig:com_Wasym_lasym}. In particular, we notice that at
high rapidities ($|y_{W}| > $ 1.4) there is a large difference in
the precision with which the as yet unmeasured $W$ production
asymmetry and the previously measured asymmetry from the decay leptons 
scaled to 1.0 fb$^{-1}$ of integrated luminosity are known.
The total systematic and statistical uncertainties on the $W$ production 
charge asymmetry measurement is shown in Fig.~\ref{fig:com_Wasym_uncern}
with the uncertainties coming from parton distribution functions
(PDFs) using CTEQ6M.  Since the systematic uncertainty estimates, as
summarized in Table~\ref{tab:syst_err}, 
are lower than the statistical error, a direct measurement of the $W$ charge 
asymmetry with this method should significantly improve  parameterizations 
of the PDFs.

\begin{figure}[t]
  \begin{center}
    \includegraphics[width=6.5cm,]{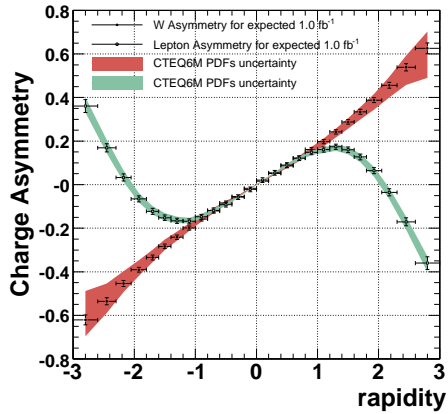}
    \caption{Comparison of statistical uncertainties expected from this
        analysis in 1.0 fb$^{-1}$ with those from the uncertainties from
        CTEQ6M PDFs~\cite{cteq6m}. \label{fig:com_Wasym_lasym}}
  \end{center}
\end{figure}

\begin{figure}[ht]
  \begin{center}
    \includegraphics[width=6.5cm,]{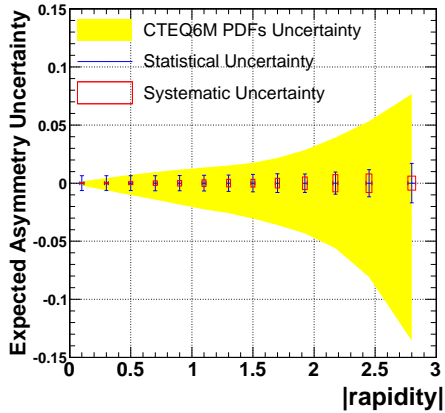}
    \caption{Total systematic uncertainty estimate and
        statistical uncertainty of this method compared
        with the current uncertainty on the $W$ charge asymmetry from
        the CTEQ6 PDFs. \label{fig:com_Wasym_uncern}}
  \end{center}
\end{figure}
In this paper, we present a study of the $W$ boson production charge
asymmetry with the $W$ decaying leptonically to an electron and
neutrino in $p\bar{p}$ collisions at the Tevatron. We propose a new
analysis technique which resolves the ambiguity in the neutrino
longitudinal momentum, using a realistic Monte Carlo simulation.  We
show that the $W$ charge asymmetry can be directly measured at the
Tevatron.  We conclude that by measuring the $W$ production charge
asymmetry with reconstructed $W$ rapidity, the result should be one of
the best determinations of the proton $d/u$ momentum ratio, and play
an important role in global PDF fits.

\bibliography{WAsym_prdrc}

\begin{thebibliography}{99}
\bibitem{coord} 
{We assume a ($z$, $\phi$, $\theta$) coordinate system with the
  $z$-axis in the direction of the proton beam; $\phi$ and $\theta$
  are the azimuthal and polar angle respectively. The pseudorapidity
  is defined as $\eta = -ln(\tan{\theta \over 2})$, and the transverse
  momentum and energy as $p_T = p \sin \theta$ and $E_T = E \sin
  \theta$, respectively.  Missing transverse energy ($\met =
  |\metvec|$) is defined as $\metvec = - \sum_{i} E_T^i
  {\bf{\hat{n}_i}}$ where ${\bf{\hat{n}_i}}$ is a unit vector in the
  transverse plane that points from the beam-line to the $i^{th}$
  calorimeter tower.}
\end{thebibliography}

\end{document}